\documentclass[12pt]{article}
\usepackage{amssymb, amsmath, amsfonts}

\newtheorem{theorem}{Theorem}

\newtheorem{remark}{Remark}

\begin{document}

\begin{center}
{\Large  On the discretization of Darboux Integrable Systems }

\bigskip

{Kostyantyn Zheltukhin}\footnote{e-mail: zheltukh@metu.edu.tr}\\

{Department of Mathematics, Middle East Technical University, Ankara, Turkey}\\

\bigskip

{Natalya Zheltukhina}\footnote{e-mail: natalya@fen.bilkent.edu.tr}\\

{Department of Mathematics, Faculty of Science, Bilkent University, Ankara, Turkey}

\end{center}

\noindent {\bf Abstract} {\it  We study the discretization of Darboux integrable systems. The discretization is done using $x$-, $y$-integrals of the considered continuous systems.
New examples of semi-discrete Darboux integrable systems are obtained.}

\noindent {\bf Keywords:} semi-discrete system, Darboux integrability, $x$-integral, $n$-integral.

\section{Introduction}

The  classification problem of Darboux integrable equations has attracted a considerable interest in the recent time, see the survey paper \cite{ZhMHSh} and references there in. There are  many   classification results in the continuous case. The case  of semi-discrete and discrete equations is not that well studied.
Discrete models play a big role in many areas of physics and discretization of existing integrable continuous models is an important problem. There is a currently discussed conjecture saying  that for each continuous Darboux integrable system it is possible to find  a semi-discrete Darboux integrable system that admits the same set of $x$-integrals. To better understand properties of semi-discrete and discrete  Darboux integrable systems it is important to have enough examples of such systems. We can test the conjecture and obtain new semi-discrete Darboux integrable systems, corresponding to  given continuous ones, following an approach  proposed by Habibullin et al., see \cite{HZhS}. In this case  we take  a Darboux  integrable continuous equation and  look for a semi-discrete equation admitting the same integrals. The method was successfully applied to many  Darboux integrable continuous equations, see \cite{HZhS}-\cite{ZhZh}. In almost all considered cases   such semi-discrete equations exist and they are Darboux integrable.

In the present paper we apply this method of discretization  to Darboux  integrable systems to obtain new Darboux integrable semi-discrete systems.
Let us give necessary definitions and formulate the main results of our work.

Consider a hyperbolic  continuous system
\begin{equation}\label{gen_cont_system}
 p_{xy}={\cal A}(p,p_x,p_y) \qquad \big( p^i_{xy}={\cal A}^i(p^1\dots p^N,p^1_x\dots p^N_x,p^1_y\dots p^N_y) \quad i=1,\dots N \big),
\end{equation}
where    $p^i(x,y)$, $i=1,\dots N$, are  functions of continuous variables $x,y\in \mathbb R$.
We say that a function $F(x,y,p,p_y, p_{yy},\dots)$ is  an $x$-integral of the system \eqref{gen_cont_system} if
\begin{equation*}
D_x F(x,y,p,p_y,p_{yy},...)=0 \qquad \mbox {on all the solutions of the system \eqref{gen_cont_system}.}
\end{equation*}
 The operator $D_x$ represents the total derivative with respect to $x$. The $y$-integral of the system \eqref{gen_cont_system} is
defined in a similar way.
The system \eqref{gen_cont_system} is called Darboux integrable if it admits $N$ functionally independent non-trivial  $x$-integrals and $N$ functionally independent non-trivial $y$-integrals.

Consider a  hyperbolic semi-discrete system
\begin{equation}\label{gen_sem-discr_system}
 q_{x1}={\cal B}(q,q_x,q_1), \qquad \big( q^i_{x1}={\cal B}^i(q^1\dots q^N,q^1_x\dots q^N_x,q^1_1\dots q_1^N)\quad i=1,\dots N \big),
\end{equation}
where $q^i(x,n)$, $i=1,\dots N$, are  functions of a continuous variable $x\in \mathbb R$ and  a discrete variable $n\in \mathbb N$. Note that we use notation  $q_1(x,n)=Dq(x,n)=q(x,n+1)$ and $q_k(x,n)=D^kq(x,n)=q(x,n+k)$,
 where $D$ is the shift operator. To state the Darboux integrability of a semi-discrete system we need to define $x$- and $n$-integrals for such systems, see \cite{HP}.
An $x$-integral is defined in the same way as in continuous case and  a function $I(x,n,q,q_x,q_{xx},\dots)$ is an $n$-integral of system \eqref{gen_sem-discr_system} if
\begin{equation*}
D I(x,n,q,q_x,q_{xx},...)=I(x,n,q,q_x,q_{xx},...) \qquad \mbox {on all the solutions of the system \eqref{gen_sem-discr_system}.}
\end{equation*}
The system \eqref{gen_sem-discr_system} is called Darboux integrable if it admits $N$ functionally independent non-trivial  $x$-integrals and $N$ functionally independent non-trivial $n$-integrals.

To find new  Darboux integrable semi-discrete systems we  applied the discretization method proposed in \cite{HZhS} to one of the continuous systems derived by Zhiber, Kostrigina  in \cite{KoZh} and continuous systems  derived by Shabat,  Yamilov in \cite{ShY}.
In \cite{KoZh}  the authors considered the classification problem for Darboux integrable continuous systems that admit the $x$- and $y$-integrals of the first and second order. In \cite{ShY}  the authors considered the exponential type system
\begin{equation*}
\mu^i_{xy}=e^{\sum a_{ij}\mu^j}, \qquad i, j=1, 2, \ldots, N.
\end{equation*}
 It was shown  that such a system is Darboux integrable if and only if the matrix $A=(a_{ij})$ is a Cartan matrix of a semi-simple Lie algebra. Such systems are closely related to the classical Toda field theories, see \cite{I}-\cite{LS} and references there in. In this case we obtain the Darboux integrable semi-discrete systems that were already described in \cite{HZhY}.

First we consider the following system (see \cite{KoZh})
\begin{equation}\label{Zh_Ko_system}
\displaystyle{
\left \{
\begin{array}{l}
u_{xy}= \frac{u_xu_y}{u+v+c}+\left(\frac{1}{u+v+c} +\frac{1}{u+v-c}  \right)u_xv_y\\
v_{xy}=\frac{v_xv_y}{u+v-c}+\left(\frac{1}{u+v+c} +\frac{1}{u+v-c}  \right)u_xv_y\, ,\\
\end{array}
\right.
}
\end{equation}
where $c$ is an arbitrary constant.
This system is Darboux integrable and admits the following $y$-integrals
\begin{equation}\label{Int1}
I_1=2v-\frac{v_x(u+v+c)}{u_x} +2c\ln\frac{u_x}{u+v+c}
\end{equation}
and
\begin{equation}\label{Int2}
I_2=\frac{u_{xx}}{u_x} - \frac{2u_x+v_x}{u+v+c}.
\end{equation}
The $x$- integrals have the same form in $u, \, v,\, u_y,\, v_y, \dots$ variables.

Now we look for  semi-discrete systems admitting these functions as $n$-integrals. The obtained results are given in  Theorems \ref{th1} and  \ref{th2} below.

\begin{theorem}\label{th1} The system
\begin{equation}\label{system}
\left \{
\begin{array}{l}
u_{1x}=f(x,n,u,v,u_1,v_1,u_x,v_x) \\
v_{1x}=g(x,n,u,v,u_1,v_1,u_x,v_x)
\end{array}
\right.
\end{equation}
possessing $n$-integrals \eqref{Int1} and \eqref{Int2}, where $c$ is a function of $n$ satisfying $c(n)\ne c(n+1)$ for all $n\in \mathbb Z$, has the form
\begin{equation}\label{TH1system}
\left \{
\begin{array}{l}
\displaystyle{u_{1x}=\frac{(u_1+v_1+c_1)u_x}{u+v+c}} \\
\\
\displaystyle{v_{1x}=\frac{2(v_1-v)u_x}{u+v+c}+\frac{2(c_1-c)u_x}{u+v+c}\ln\frac{u_x}{u+v+c}+v_x}\, .
\end{array}
\right.
\end{equation}
Moreover, the system above also possesses $x$-integrals
\begin{equation}\label{TH1F_1}
F_1=\frac{(c-c_1)(v_2-v)-(c-c_2)(v_1-v)}{(c-c_2)(v_3-v)-(c-c_3)(v_2-v)}
\end{equation}
and
\begin{equation}\label{TH1F_2}
F_2=\frac{(c_1-c_2)u+(c_2-c)u_1+(c-c_1)u_2}{\sqrt{(c_1-c_2)v+(c_2-c)v_1+(c-c_1)v_2}}-\sqrt{(c_1-c_2)v+(c_2-c)v_1+(c-c_1)v_2}\,
.
\end{equation}
Hence, semi-discrete system {\eqref{TH1system}} is Darboux integrable.
\end{theorem}

\begin{theorem}\label{th2} The system \eqref{system}
possessing $n$-integrals {\eqref{Int1}} and  {\eqref{Int2}},
where $c$ is a  constant, is either
\begin{equation}\label{TH2system1}
\left \{
\begin{array}{l}
\displaystyle{u_{1x}=\frac{(u_1+v_1+c)u_x}{u+v+c}} \\
\\
\displaystyle{v_{1x}=\frac{2(v_1-v)u_x}{u+v+c}+v_x}
\end{array}
\right.
\end{equation}
with $x$-integrals $\displaystyle{F_1=\frac{v_1-v}{v_2-v_1}}$ and $\displaystyle{F_2=\frac{u_2-u+v-v_2}{\sqrt{v_1-v}}}$,
or
\begin{equation}\label{TH2system2}
\left \{
\begin{array}{l}
\displaystyle{u_{1x}=\frac{(u_1+v_1+c)Bu_x}{u+v+c}} \\
\\
\displaystyle{v_{1x}=\frac{2B(v_1-v+c\ln B)}{u+v+c}u_x+Bv_x}\, ,
\end{array}
\right.
\end{equation}
where $B$ is defined by equality $H(K_1, K_2)=0$ with
$$
 K_1=\frac{v_1-vB+B(1-B)u+c\ln B}{(B-1)^2}+c\ln (B-1)-c\ln B
 $$
 and $$
 K_2=\frac{u_1+cB-c-c\ln B}{B-1}+\frac{B^2v-Bv_1-cB\ln B}{(B-1)^2}+c\ln (B-1)-c\ln B,
 $$
and
$H$ being any smooth function.
\end{theorem}

\begin{remark}

We considered some special cases of the system  \eqref{TH2system2} and get  Darboux integrable systems.

\noindent (I)  System {\eqref{TH2system2}} with $B=\displaystyle{\frac{u-v+(-1)^n\sqrt{(u-v)^2+4uv_1}}{2u}}$ is Darboux Integrable. (The expression for $B$ is found from $K_1=0$, with $c=0$.)

\bigskip

\noindent (II) System {\eqref{TH2system2}} with $B=\displaystyle{\frac{v_1-u_1+(-1)^n\sqrt{(v_1-u_1)^2+4u_1v}}{2v}}$ is Darboux Integrable. (The expression for $B$ is found from $K_2=0$, with $c=0$.)
\end{remark}

\begin{remark}
Expansion of the function $B(u,v,v_1)$, given implicitly by $(B-1)^2K_1=0$, into a series of the form
\begin{equation}
B(u,v,v_1)= \sum\limits_{n=0}^\infty a_n(v_1-v)^n,
\end{equation}
where coefficients $a_n$ depend on variables $u$ and $v$, yields $a_0=1$ and $a_1=\dfrac{1}{u+v-c}$. So $B$ can be written as
\begin{equation}
B(u,v,v_1)= 1+\frac{1}{u+v-c}(v_1-v)+\sum\limits_{n=2}^\infty a_n(v_1-v)^n.
\end{equation}
By letting $u_1=u+\epsilon u_y$ and  $v=v+\epsilon v_y$ and taking $\epsilon\to 0$ one can see that the system \eqref{TH2system2} has a continuum limit \eqref{Zh_Ko_system}.
\end{remark}

Let us discuss the  exponential type systems.
We consider the discretization  of such systems corresponding to $2\times 2$ matrices, namely,
\begin{equation}\label{cont_exp_system}
\begin{array}{l}
\mu_{xy}=e^{2\mu-\nu},\\
\nu_{xy}=e^{-c\mu-2\nu},
\end{array}
\end{equation}
where $c=1,2,3$.
The obtained results are given in  Theorem \ref{th3} below.
The discretization of such systems was also considered in \cite{HZhY}, where the form of the corresponding semi-discrete system was directly postulated and then the Darboux integrability proved. In our approach we do not make any specific assumptions  about the form of the corresponding semi-discrete system.
Note that the integrals corresponding to Darboux integrable exponential systems are given in the statement of  Theorem \ref{th3}.
\begin{theorem}\label{th3}
\noindent
(1) The system
\begin{equation}\label{systemCartan}
\left \{
\begin{array}{l}
u_{1x}=\tilde f(u,v,u_1,v_1,u_x,v_x) \\
v_{1x}=\tilde g(u,v,u_1,v_1,u_x,v_x),
\end{array}
\right.
\end{equation}
possessing $n$-integrals
\begin{equation}\label{I_1}
I_1=u_{xx}+v_{xx}-u_x^2+u_xv_x-v_x^2
\end{equation}
and
\begin{equation}\label{I_1^*}
I_1^*=u_{xxx}+u_x(v_{xx}-2u_{xx})+u_x^2v_x-u_xv_x^2
\end{equation}
has the form
\begin{equation}\label{Examnple4_1}
\left\{
\begin{array}{lll}
u_{1x} & = &u_x + Ae^{u_1+u-v_1}\\
v_{1x} & = & v_x + Be^{-u+v+v_1}\, , \\
\end{array}\right.
\end{equation}
or
\begin{equation}\label{Example4_2}
\left\{
\begin{array}{lll}
u_{1x} & = &u_x + Ae^{u_1+u-v}\\
v_{1x} & = & v_x + Be^{-u_1+v+v_1}\, , \\
\end{array}\right.
\end{equation}
where $A$ and $B$   are arbitrary constants.\\
(2)   The system \eqref{systemCartan} possessing $n$-integrals
\begin{equation}\label{I_2}
I_2=2u_{xx}+v_{xx}-2u_x^2+2u_xv_x-v_x^2
\end{equation}
and
\begin{multline}\label{I_2^*}
I_2^*=u_{xxxx}+u_x(v_{xxx}-2u_{xxx})+u_{xx}(4u_xv_x-2u_x^2 -v_x^2)\\
+u_{xx}(v_{xx}-u_{xx}) + v_{xx}u_x(u_x-2v_x) +u_x^4+u_x^2v_x^2-2u_x^3v_x
\end{multline}
has the form \begin{equation}\left\{
\begin{array}{lll}\label{Example5_1}
u_{1x} & = &u_x + Ae^{u+u_1-v_1}  \\
v_{1x} & = &v_x + Be^{-2u+v+v_1}\, ,
\end{array}\right.
\end{equation}
where $A$ and $B$   are arbitrary constants.\\
(3)  The system \eqref{systemCartan} possessing $n$-integrals
\begin{equation}\label{I_3}
I_3=u_{xx}+\frac{1}{3}v_{xx}-u_x^2+u_xv_x-\frac{1}{3}v_x^2
\end{equation}
and
\begin{multline}\label{I_3^*}
I_3^*=u_{(6)} - 2u_{(5)}u_x + v_{(5)}u_x + u_{(4)}(32(u_x)^2 - 30u_xv_x + 11(v_x)^2 - 40u_{xx} - 11v_{xx}) \\
+ v_{(4)}(14(u_x)^2 - 15u_xv_x + (13/3)(v_x)^2 - 10u_{xx} - (13/3)v_{xx} ) +19(u_{(3)})^2 + (13/6)(v_{(3)})^2 + 16u_{(3)}v_{(3)}\\
 + u_{(3)}(-36u_{xx}u_x + 18u_{xx}v_x + 80v_{xx}u_x - 45v_{xx}v_x) +v_{(3)}(-52u_{xx}u_x + 33u_{xx}v_x - 5v_{xx}u_x) \\
 + u_{(3)}(-64(u_x)^3 + 102(u_x)^2v_x - 62u_x(v_x)^2 + 13(v_x)^3) + v_{(3)}(32(u_x)^3 - 58(u_x)^2v_x \\
 + 38u_x(v_x)^2 - (26/3)(v_x)^3) + 66(u_{xx})^3 + (26/3)(v_{xx})^3 - 35(u_{xx})^2(v_{xx}) - 5u_{xx}(v_{xx})^2 \\
 +(u_{xx})^2(30(u_x)^2 - 18u_xv_x - (11/2)(v_x)^2) + u_{xx}v_{xx}(-34(u_x)^2 + 32u_xv_x - 2(v_x)^2) - 2(v_{xx})^2u_xv_x \\
 +u_{xx}(6(u_x)^4 - 24(u_x)^3v_x + 25(u_x)^2(v_x)^2 - 9u_x(v_x)^3 + (v_x)^4) + v_{xx}(-(u_x)^4 + 8(u_x)^3v_x - 8(u_x)^2(v_x)^2 \\
 + 2u_x(v_x)^3) + (-2(u_x)^6 + 6(u_x)^5v_x - (13/2)(u_x)^4(v_x)^2 + 3(u_x)^3(v_x)^3 - (1/2)(u_x)^2(v_x)^4)
\end{multline}
has the form
\begin{equation}\left\{
\begin{array}{lll}\label{Example6}
u_{1x} & = & u_x+Ae^{u+u_1-v_1}  \\
v_{1x} & = & v_x+Be^{-3u+v+v_1}\, ,
\end{array}\right.
\end{equation}
where $A$ and $B$ are arbitrary constants.
\end{theorem}

\begin{remark}
We note that while considering systems with integrals \eqref{I_2} and \eqref{I_2^*}  we also obtain two degenerate systems
\begin{equation}\left\{
\begin{array}{lll}\label{Example5_2}
u_{1x} & = &u_x  \\
v_{1x} & = &v_x + Be^{-(2+c)u+cu_1+v+v_1}\, ,
\end{array}\right.
\end{equation}
and
\begin{equation}\left\{
\begin{array}{lll}\label{Example5_3}
u_{1x} & = &u_x + Ae^{u+u_1+2cv-(2c+1)v_1}  \\
v_{1x} & = &v_x \, ,
\end{array}\right.
\end{equation}
where $A$, $B$  and $c$ are arbitrary constants,which are equivalent to a Darboux integrable equation.
\end{remark}
\begin{remark}
By letting $u=\mu^1$, $u_1=\mu^1+\epsilon \mu^1_y$, $v=\mu^2$,  $v_1=\mu^2+\epsilon \mu^2_y$  and $A=\epsilon$, $B=\epsilon$ in equations  \eqref{Examnple4_1}, \eqref{Example5_1}, \eqref{Example6} and taking $\epsilon\to 0$ one can see that the considered systems have corresponding continuum limit given by \eqref{cont_exp_system}.
\end{remark}

\section{Proof of Theorems 1.1 and 1.2}

Let us find a semi-discrete system {\eqref{system}} possessing $n$-integrals {\eqref{Int1}} and  {\eqref{Int2}},
where $c$ is an arbitrary constant, possibly dependent on $n$. Let $Dc=c_1$.
It follows from $DI_2=I_2$ that
\begin{equation*}
\frac{u_{1xx}}{u_{1x}} - \frac{2u_{1x}+v_{1x}}{u_1+v_1+c_1}=\frac{u_{xx}}{u_x} - \frac{2u_x+v_x}{u+v+c},
\end{equation*}
that is
\begin{equation}\label{Int2_eq}
\frac{f_x+f_uu_x+f_vv_x+f_{u_1}f+f_{v_1}g +f_{u_x}u_{xx}+f_{v_x}v_{xx}}{f}- \frac{2f+g}{u_1+v_1+c_1}=\frac{u_{xx}}{u_x} - \frac{2u_x+v_x}{u+v+c}\, .
\end{equation}
Compare the coefficients by $v_{xx}$ and $u_{xx}$, we get $f_{v_x}=0$ and $\dfrac{f_{u_x}}{f}=\dfrac{1}{u_x}$. Hence
\begin{equation}\label{new_f}
f(x,n,u,v,u_1,v_1,u_x,v_x)=A(x,n,u,v,u_1,v_1)u_x.
\end{equation}
It follows from $DI_1=I_1$ that
\begin{equation}
2v_1-\frac{(u_1+v_1+c_1)g}{f} +2c_1\ln\frac{f}{u_1+v_1+c_1}= 2v-\frac{v_x(u+v+c)}{u_x} +2c\ln\frac{u_x}{u+v+c}.
\end{equation}
Using \eqref{new_f} we obtain
\begin{equation*}
2v_1-\frac{(u_1+v_1+c_1)g}{Au_x} +2c_1\ln\frac{Au_x}{u_1+v_1+c_1}= 2v-\frac{v_x(u+v+c)}{u_x} +2c\ln\frac{u_x}{u+v+c}
\end{equation*}
and find $g$ as
\begin{multline}\label{new_g}
g=\left(\frac{2(v_1-v)A}{(u_1+v_1+c_1)} +\frac{2Ac_1}{(u_1+v_1+c_1)}\ln\frac{(u+v+c)A}{(u_1+v_1+c_1)}\right)u_x+
\frac{2(c_1-c)A}{(u_1+v_1+c_1)}u_x\ln\frac{u_x}{u+v+c} \\
+\frac{(u+v+c)A}{(u_1+v_1+c_1)}v_x\, .
\end{multline}
Substituting the expressions \eqref{new_f} and \eqref{new_g} into equality \eqref{Int2_eq}  and comparing coefficients by $u_x$, $v_x$, $u_x\ln\dfrac{u_x}{u+v+c}$
 and free term we get the following equalities
\begin{multline}\label{eq_A1}
\shoveright{\frac{A_x}{A} =0}
\end{multline}
\begin{multline}\label{eq_A2}
\shoveright{\frac{2(c_1-c)A_{v_1}}{(u_1+v_1+c_1)}-\frac{2(c_1-c)A}{(u_1+v_1+c_1)^2}=0}
\end{multline}
\begin{multline}\label{eq_A3}
 \frac{A_u}{A}+A_{u_1}+\left(  \frac{A_{v_1}}{A}- \frac{1}{(u_1+v_1+c_1)}\right)\left( \frac{2(v_1-v)A}{(u_1+v_1+c_1)}+\frac{2c_1A}{(u_1+v_1+c_1)}\ln\frac{(u+v+c)A}{(u_1+v_1+c_1)}\right)\\
 - \frac{2A}{(u_1+v_1+c_1)} + \frac{2}{(u+v+c)}=0
\end{multline}
\begin{multline}\label{eq_A4}
 \shoveright{\frac{A_v}{A}+\frac{(u+v+c)A_{v_1}}{(u_1+v_1+c_1)}-\frac{(u+v+c)A}{(u_1+v_1+c_1)^2}+\frac{1}{(u+v+c)}=0\, .}
\end{multline}
We have two possibilities: $c_1\ne c$ and $c_1=c$.

\subsection{$c$ depends on $n$}

First we consider the case $c_1\ne c$, that is $c$ depends on $n$ and satisfies $c(n)\ne c(n+1)$ for all $n$. Then equations \eqref{eq_A2}-\eqref{eq_A4} are transformed into
\begin{multline}\label{eq_newA2}
\shoveright{\frac{A_{v_1}}{A}-\frac{1}{(u_1+v_1+c_1)}=0}
\end{multline}
\begin{multline}\label{eq_newA3}
 \shoveright{\frac{A_u}{A}+A_{u_1}- \frac{2A}{(u_1+v_1+c_1)}+\frac{2}{(u+v+c)}=0}
\end{multline}
\begin{multline}\label{eq_newA4}
\shoveright{ \frac{A_v}{A}+\frac{1}{(u+v+c)}=0\, .}
\end{multline}
Equations \eqref{eq_newA2} and \eqref{eq_newA4} imply that
\begin{equation}
A=\frac{(u_1+v_1+c_1)}{(u+v+c)}M(n,u,u_1).
\end{equation}
Substituting the above $A$ into \eqref{eq_newA3} we get that $M$ satisfies
\begin{equation}\label{eq_M}
(u+v+c)\frac{M_u}{M} + (u_1+v_1+c_1)M_{u_1}+(1-M)=0.
\end{equation}
Differentiating equation \eqref{eq_M} with respect to $v$ and $v_1$ we get that $M_u=0$ and $M_{u_1}=0$ respectively. Thus,  equation \eqref{eq_M} implies that $M=1$.
So in the case  $c_1\ne c$ we arrive to the  system of equations \eqref{TH1system}.
We note that the system  \eqref{TH1system} is Darboux integrable. It admits two $n$-integrals \eqref{Int1} and \eqref{Int2} and two $x$-integrals
\eqref{TH1F_1} and \eqref{TH1F_2}.
The $x$-integrals can be found by considering the characteristic $x$-ring for system \eqref{TH1system}.

\subsection{$c$ does not  depend on $n$ }

Now we consider the case $c=c_1$, that is $c$ is a constant independent of $n$. Then we have equations \eqref{eq_A3} and \eqref{eq_A4}.
Introducing new variable $B=\dfrac{(u+v+c)}{(u_1+v_1+c)}A$ we can rewrite the equations as
\begin{multline}\label{eq_B1}
 \shoveright{\frac{B_u}{B}+\frac{(u_1+v_1+c)}{(u+v+c)}B_{u_1} +2\frac{(v_1-v+c\ln B)}{(u+v+c)}B_{v_1}+ \frac{1-B}{(u+v+c)}=0}
\end{multline}
\begin{multline}\label{eq_B2}
\shoveright{ \frac{B_v}{B}+B_{v_1}=0\, .}
\end{multline}
 The set of solutions of the above system is not empty, for example it admits a solution $B=1$. Setting $B=1$ we  arrive to the  system of equations \eqref{TH2system1}.
We note that the system  \eqref{TH2system1} is Darboux integrable. It admits two $n$-integrals \eqref{Int1} and \eqref{Int2} and two $x$-integrals
\begin{equation*}
F_1=\frac{v_1-v}{v_2-v_1}\, ,\qquad
F_2=\frac{u_2-u+v-v_2}{\sqrt{v_1-v}}.
\end{equation*}
The $x$-integrals are calculated by considering the characteristic $x$-ring for system \eqref{TH2system1}.

\noindent Now let us consider case when $B\ne 1$ identically. For function $W=W(u,v, u_1, v_1, B)$ equations \eqref{eq_B1} and \eqref{eq_B2} become
\begin{multline}\label{eq_W1}
 \shoveright{\frac{W_u}{B}+\frac{(u_1+v_1+c)}{(u+v+c)}W_{u_1} +2\frac{(v_1-v+c\ln B)}{(u+v+c)}W_{v_1}+ \frac{B-1}{(u+v+c)}W_B=0}
\end{multline}
\begin{multline}\label{eq_W2}
\shoveright{ \frac{W_v}{B}+W_{v_1}=0\, .}
\end{multline}
After the change of variables $\tilde{v}=v+c$, $\tilde{v_1}=v_1+c-(v+c)B$, $\tilde{u}=u$, $\tilde{u_1}=u_1$, $\tilde{B}=B$ equations  \eqref{eq_W2} and \eqref{eq_W1} become
$W_{\tilde{v}}=0$ and
$$
\frac{\tilde{u}+\tilde{v}}{\tilde{B}}W_{\tilde{u}}+(\tilde{u_1}+\tilde{v_1}+\tilde{v}\tilde{B})W_{\tilde{u_1}}+(2\tilde{v_1}+2c\ln\tilde{B}+\tilde{v}(\tilde{B}-1))W_{\tilde{v_1}}+(\tilde{B}-1)W_{\tilde{B}}=0.
$$
We differentiate the last equality with respect to $\tilde{v}$, use $W_{\tilde{v}}=0$, and find that $W$ satisfies the following equations
\begin{equation*}
 \shoveright{\frac{W_{\tilde{u}}}{\tilde{B}}+\tilde{B}W_{\tilde{u_1}}+(\tilde{B}-1)W_{\tilde{v_1}}=0}
\end{equation*}
\begin{equation*}
\shoveright{ \frac{\tilde{u}}{\tilde{B}}W_{\tilde{u}}+(\tilde{u_1}+\tilde{v_1})W_{\tilde{u_1}}+(2\tilde{v_1}+2c\ln\tilde{B})W_{\tilde{v_1}}+(\tilde{B}-1)W_{\tilde{B}}=0\, .}
\end{equation*}
After doing another change of variables $u_1^*=\tilde{u_1}-\tilde{B}^2\tilde{u}$, $v_1^*=\tilde{v_1}+\tilde{B}(1-\tilde{B})\tilde{u}$, $u^*=\tilde{u}$, $B^*=\tilde{B}$, we obtain that
$W_{u^*}=0$ and
\begin{equation*}
 \shoveright{(u_1^*+v_1^*)W_{u_1^*}+(2v_1^*+2c\ln B^*)W_{v_1^*}+(B^*-1)W_{B^*}=0.}
 \end{equation*}
 The first integrals of the last equation are
 $$
 K_1=\frac{v_1^*}{(B^*-1)^2}+\frac{c\ln B^*}{(B^*-1)^2}-c\ln B^*+c\ln (B^*-1)+\frac{c}{B^*-1}
 $$
 and
 $$
 K_2=\frac{u_1^*-c-c\ln B^*}{B^*-1}-\frac{B^*v^*_1}{(B^*-1)^2}-\frac{cB^*\ln B^*}{(B^*-1)^2}+c\ln (B^*-1)-c\ln B^*.
 $$
 They can be rewritten in the original variables as
 $$
 K_1=\frac{v_1-vB+B(1-B)u+c\ln B}{(B-1)^2}+c\ln (B-1)-c\ln B
 $$
 and $$
 K_2=\frac{u_1+cB-c-c\ln B}{B-1}+\frac{B^2v-Bv_1-cB\ln B}{(B-1)^2}+c\ln (B-1)-c\ln B.
 $$
Therefore, system \eqref{system} becomes \eqref{TH2system2} due to \eqref{new_f} and \eqref{new_g}.

 \subsection {Proof of  Remark 1.1}

 Function $B$ is any function satisfying the equality $H(K_1, K_2)=0$, where $H$ is any smooth function.

 \noindent (I) By taking function $H$ as  $H(K_1,K_2)=K_1$ we obtain  one possible  function $B$. It  satisfies the
 equality $-uB^2+(u-v)B+v_1=0$ and can be taken as
$B=\displaystyle{\frac{u-v+(-1)^n\sqrt{(u-v)^2+4uv_1}}{2u}}$

\noindent (II) By taking function $H$ as  $H(K_1,K_2)=K_2$ we obtain  another possible  function $B$. It  satisfies the
equality  $vB^2+(u_1-v_1)B-u_1=0$ and can be taken as
  $B=\displaystyle{\frac{v_1-u_1+(-1)^n\sqrt{(v_1-u_1)^2+4u_1v}}{2v}}$.

  In both cases ((I) and (II)) let us consider the corresponding $x$-rings. Denote by $X=D_x$, $Y_1=\displaystyle{\frac{\partial}{\partial u_x}}$,
  $Y_2=\displaystyle{\frac{\partial}{\partial v_x}}$, $E_1=\displaystyle{\frac{u+v}{B}[Y_1,X]}$, $E_2=\displaystyle{\frac{1}{B}[Y_2,X]}$, $E_3=[E_1, E_2]$.
  Note that $X=u_xE_1+v_xE_2$. We have,
  $$\begin{array}{l|ccc}
[E_i, E_j]& E_1&E_2&E_3\\
\hline
E_1&0&E_3&\alpha_1E_2+\alpha_2E_3\\
E_2&-E_3&0&0\\
E_3&-(\alpha_1E_2+\alpha_2E_3)&0&0
\end{array}
$$
where
$$\alpha_1 =\frac{2v_1(u-v)+2(uv-v^2+2uv_1)B}{v_1(u-v)+((u-v)^2+2uv_1)B}\, , \qquad \alpha_2=-3+\frac{2}{B}
$$
in case (I) and
$$
\alpha_1= \frac{2u_1^2+4u_1v-2u_1v_1+2(-(u_1-v_1)^2+vv_1-3vu_1)B}{u_1(v_1-u_1)+((u_1-v_1)^2+2u_1v)B}  \,,  \qquad \alpha_2=-3+\frac{2}{B}
$$
in case (II).

\section{Proof of Theorem 1.3}
\subsection{Case (1)}
Let us find a system
\begin{equation}\label{ex4_system_0}\left\{
\begin{array}{lll}
u_{1x} &=& \tilde f(x,n,u,v,u_1, v_1, u_x,v_x)   \\
v_{1x} &=& \tilde g (x,n,u,v,u_1, v_1, u_x,v_x)
\end{array}\right.
\end{equation}
possessing $n$-integrals \eqref{I_1} and \eqref{I_1^*}. The equality $DI=I$ implies
\begin{multline}
u_{1xx}+v_{1xx}-u_{1x}^2+u_{1x}v_{1x}-v_{1x}^2=u_{xx}+v_{xx}-u_x^2+u_xv_x-v_x^2,
\end{multline}
or the same
\begin{multline}\label{ex4_int_I}
\tilde f_x+\tilde f_uu_x+\tilde f_vv_x+\tilde f_{u_1}\tilde f+\tilde f_{v_1}\tilde g+\tilde f_{u_x}u_{xx}+\tilde f_{v_x}v_{xx}+
\tilde g_x+\tilde g_uu_x+\tilde g_vv_x \\+\tilde g_{u_1}\tilde f+ \tilde g_{v_1}\tilde g+\tilde g_{u_x}u_{xx}+ \tilde g_{v_x}v_{xx}- \tilde f^2+\tilde f\tilde g-\tilde g^2=
u_{xx}+v_{xx}-u_x^2+u_xv_x-v_x^2\, .
\end{multline}
We consider the coefficients by $u_{xx}$ and $u_{xx}$ in \eqref{ex4_int_I} to get
\begin{eqnarray}
\tilde f_{u_x}+\tilde g_{u_x} &=&1  \label{ex4_1} \\
\tilde f_{v_x}+\tilde g_{v_x} &=&1. \label{ex4_2}
\end{eqnarray}
The equality $DI_1^*=I_1^*$ implies
\begin{equation}\label{ex4_int_I*}
u_{1xxx}+u_{1x}(v_{1xx}-2u_{1xx})+u_{1x}^2v_{1x}-u_{1x}v_{1x}^2=
u_{xxx}+u_x(v_{xx}-2u_{xx})+u_x^2v_x-u_xv_x^2.
\end{equation}
Since $DI_1^*=u_{1xxx}+\dots=\tilde f_{u_x}u_{xxx}+\dots $, where the remaining terms do not depend on $u_{xxx}$, the equality \eqref{ex4_int_I*} implies
\begin{equation} \label{ex4_3}
\tilde f_{u_x}=1.
\end{equation}
Note that $J=D_xI_1-I_1^*=v_{xxx}+v_x(u_{xx}-2v_{xx})+v_x^2u_x-u_x^2v_x$ is an $n$-integral as well.  Since $DJ=J$ and $DJ=v_{1xxx}+\dots=\tilde g_{v_x}v_{xxx}+\dots $, where the remaining terms do not depend on $v_{xxx}$, then
\begin{equation}\label{ex4_4}
\tilde g_{v_x}=1.
\end{equation}
It follows from equalities \eqref{ex4_1}, \eqref{ex4_2}, \eqref{ex4_3} and \eqref{ex4_4} that $\tilde f_{v_x}=0$ and $\tilde g_{u_x}=0$. Therefore the system \eqref{ex4_system_0} and equality \eqref{ex4_int_I} become
\begin{equation}\label{ex4_system_1}\left\{
\begin{array}{lll}
u_{1x} &=& u_x + f(x,n,u,v,u_1,v_1)   \\
v_{1x} &=& v_x + g(x,n,u,v,u_1,v_1)
\end{array}\right.
\end{equation}
and
\begin{multline}\label{ex4_int_I+}
 f_x + f_uu_x + f_vv_x + f_{u_1}(u_x + f) + f_{v_1}(v_x + g) + g_x + g_uu_x + g_vv_x + g_{u_1}(u_x + f)\\
 + g_{v_1}(v_x + g) - 2u_xf - f^2 +u_xg + v_xf +fg-2v_xg-g^2=0
\end{multline}
By considering  coefficients by $u_x$, $v_x$ and $u_x^0v_x^0$ in the last equality, we get
\begin{eqnarray}
(f+g)_u + (f+g)_{u_1} + (f+g) - 3f & = &0  \label{ex4_equality_**}\\
(f+g)_v + (f+g)_{v_1} + (f+g) - 3g & = &0 \label{ex4_equality_***}\\
f(f+g)_{u_1} + g(f+g)_{v_1} + (f+g)_x - (f+g)^2 + 3fg & = & 0 \, . \label{ex4_equality_****}
\end{eqnarray}
Now let us rewrite inequality \eqref{ex4_int_I*} for the system \eqref{ex4_system_1}
\begin{multline}\label{ex4_system_C}
D_x \big( f_x + f_uu_x + f_vv_x + f_{u_1}(u_x+f) + f_{v_1}(v_x+g) \big) \\
+ (u_x+f)\big( g_x + g_uu_x + g_vv_x + g_{u_1}(u_x+f) + g_{v_1}(v_x+g) + v_{xx} \big )\\
+ (u_x+f)\big( -2f_x -2f_uu_x - 2f_vv_x -2 f_{u_1}(u_x+f) - 2f_{v_1}(v_x+g)-2u_{xx} \big ) \\
+ (u_x^2 + 2u_xf + f^2)(v_x + g)-(v_x^2 + 2v_xg + g^2)(u_x+f)
= u_x(v_{xx}- 2 u_{xx}) + u_x^2v_x - u_xv_x^2\, .
\end{multline}
By comparing the coefficients by $u_{xx}$ and  $v_{xx}$ in the last equality, we get
\begin{equation}\label{ex4_system_A}
\begin{array}{lll}
f_u + f_{u_1}  = 2f \\
f_v + f_{v_1} = -f\, . \\
\end{array}
\end{equation}
It follows from equality $DJ=J$ that
\begin{multline}\label{ex4_system_G}
D_x \big( g_x + g_uu_x + g_vv_x + g_{u_1}(u_x+f) + g_{v_1}(v_x+g) \big) \\
+(v_x+g)\big( f_x + f_uu_x + f_vv_x + f_{u_1}(u_x+f) + f_{v_1}(v_x+g) + u_{xx} \big )\\
- 2(v_x+g)\big( g_x + g_uu_x + g_vv_x + g_{u_1}(u_x+f) + g_{v_1}(v_x+g) + v_{xx} \big ) \\
+ (u_x + f)(v_x^2 + 2v_xg + g^2)-(v_x+g)(u_x^2 + 2u_xf + f^2)
= v_x(u_{xx}- 2 v_{xx}) + v_x^2u_x - u_x^2v_x\, .
\end{multline}
By comparing the coefficients by $u_{xx}$ and  $v_{xx}$ in the last equality, we get
\begin{equation}\label{ex4_system_B}
\begin{array}{lll}
g_u + g_{u_1}  = -g \\
g_v + g_{v_1} = 2g \, .\\
\end{array}
\end{equation}

\noindent Note that the equalities \eqref{ex4_equality_**} and \eqref{ex4_equality_***} follow from equalities \eqref{ex4_system_A} and \eqref{ex4_system_B}.
Let us use equalities \eqref{ex4_system_A} and \eqref{ex4_system_B} to rewrite equality \eqref{ex4_system_C}
\begin{multline}\nonumber
D_x ( f_x + 2fu_x - fv_x + f_{u_1}f + f_{v_1}g )
+ (u_x+f)( g_x + g_{u_1}f + g_{v_1}g + v_{xx} - 4fu_x-2f_x) \\
 + (u_x+f)( 2fv_x - 2f_{u_1}f - 2f_{v_1}g-2u_{xx} + u_xv_x + fv_x + fg - v_x^2 - g^2)\\
= u_x(v_{xx}- 2 u_{xx}) + u_x^2v_x - u_xv_x^2\, .
\end{multline}
We note that  the consideration of the coefficients by $u_{xx}$, $v_{xx}$,   $u_x^2$, $v_x^2$, $u_xv_x$  in the above equality give us
equations that follow immediately from \eqref{ex4_system_A} and \eqref{ex4_system_B}. Considering coefficient by $u_x$ we get
\begin{multline}\nonumber
f_{xu} + f_{xu_1} + 2f_x + 2ff_{u_1} + 2f_{v_1}g + ff_{u_1u} + f_{u_1}f_u + f_{u_1}^2 + gf_{v_1u} \\
+gf_{u_1v_1} + f_{v_1}g_u +f_{v_1}g_{u_1}+f_{u_1u_1}f+ g_x + g_{u_1}f + g_{v_1}g - 2f_x - 2f_{u_1}f - 2f_{v_1}g + fg - g^2 - 4f^2=0.
\end{multline}
Using equations \eqref{ex4_system_A} and \eqref{ex4_system_B} we get
\begin{equation}\nonumber
2f_x+g_x+4ff_{u_1}+f_{v_1}g+g_{u_1}f+g_{v_1}g+fg-g^2-4f^2=0\, ,
\end{equation}
or using equation \eqref{ex4_equality_****} ,
\begin{equation}\label{ex4_system_D}
f_x+3f(f_{u_1}-f)=0.
\end{equation}
Considering coefficient by $v_x$ we get
\begin{multline}\nonumber
f_{xv}+f_{xv_1} - f_{x} - ff_{u_1} - f_{v_1}g + ff_{u_1v} + ff_{u_1v_1} + f_{u_1}f_v + f_{u_1}f_{v_1} \\
+gf_{v_1v} + gf_{v_1v_1}  + f_{v_1}g_v + f_{v_1}g_{v_1} + 3f^2=0.
\end{multline}
Using equations \eqref{ex4_system_A} and \eqref{ex4_system_B} we get
\begin{equation}\label{ex4_system_E}
2f_x+3f(f_{u_1}-f)=0.
\end{equation}
It follows from equations \eqref{ex4_system_D} and \eqref{ex4_system_E} that $f_x=0$ and $f(f_{u_1}-f)=0$. Thus either $f=0$ or
\begin{equation}\label{ex4_fufu1}
\left\{
\begin{array}{l}
f=f_{u_1}\\
f=f_u.
\end{array}
\right.
 \end{equation}
Now we consider the coefficient by $u_x^0v_x^0$ in \eqref{ex4_system_C} we get
\begin{multline}\nonumber
f^2f_{u_1u_1} + fgf_{u_1v_1} + ff_{u_1}^2 + f_{u_1}f_{v_1}g + fgf_{u_1v_1} + g^2f_{v_1v_1} + f_{v_1}g_x + ff_{v_1}g_{u_1} \\
+gf_{v_1}g_{v_1} + fg_x +f^2g_{u_1}  + fgg_{v_1} - 2f^2f_{u_1} - 2fgf_{v_1} + f^2g -fg^2=0.
\end{multline}
First assume that $f\ne 0$ then using \eqref{ex4_fufu1} we can rewrite the above equality as
\begin{equation}\label{ex4_system_F}
fgf_{v_1} + g^2f_{v_1v_1} + f_{v_1}g_x + f_{v_1}g_{u_1}f + f_{v_1}g_{v_1}g + fg_x + f^2g_{u_1}+ fgg_{v_1}+f^2g-fg^2=0\, .
\end{equation}
 Also we can rewrite equality \eqref{ex4_system_G}, using equations \eqref{ex4_system_A}, \eqref{ex4_system_B} and \eqref{ex4_equality_****} then  considering coefficients by $u_x$ and $v_x$ we obtain
 \begin{eqnarray*}
2g_x+3g(g_{v_1}-g)=0\\
g_x+3g(g_{v_1}-g)=0.
\end{eqnarray*}
From above equalities and \eqref{ex4_system_B} it follows that $g_x=0$, $g_{v_1}=g$ and $g_v=g$ (we assume that $g\ne 0$).
We have
\begin{equation}\label{ex4_system_H}
\begin{array}{llr}
f_{u_1}=f, & f_u=f,   & f_v+f_{v_1}=-f \\
g_{v_1}=g, & g_{v}=g, & g_u+g_{u_1}=-g \\
          &          & f_{v_1}g+g_{u_1}f=-fg\, .\\
\end{array}
\end{equation}
Using \eqref{ex4_system_H}, the equality \eqref{ex4_system_F} takes form $g_{u_1}f_{v_1}(-g+f)=0$.
This equality implies that  under assumptions that $f\neq 0$ and $g\neq 0$ we have  three possibilities:
(I)  $g_{u_1}=0$, (II)  $f_{v_1}=0$ and (III)   $g=f$.
Let us consider these possibilities.

\noindent
{\bf Case (I)} From $g_{u_1}=0$, using \eqref{ex4_system_H}, we get that $g_u=-g$, $g_{v_1}=g$, $g_v=g$. Thus $g=Be^{-u+v+v_1}$, where $B$ is a constant.
We also get that $f_u=f$, $f_{u_1}=f$, $f_v=0$ and $f_{v_1}=-f$. Thus $f=Ae^{u_1+u-v_1}$, where $A$ is a constant.
So the system \eqref{ex4_system_1} takes form \eqref{Examnple4_1}.

\noindent
{\bf Case (II)} From $f_{v_1}=0$, using \eqref{ex4_system_H}, we get that $f_u=f$, $f_{u_1}=f$, $f_v=-f$. Thus $f=Ae^{u_1+u-v}$, where $A$ is a constant.
We also get that $g_u=0$, $g_{u_1}=-g$, $g_v=g$ and $g_{v_1}=g$. Thus $g=Be^{-u_1+v_1+v}$, where $B$ is a constant.
So the system \eqref{ex4_system_1} takes form \eqref{Example4_2}.

\noindent
{\bf Case (III)} From $g=f$, using \eqref{ex4_system_H}, we get that $f=0$ and $g=0$.
So the system \eqref{ex4_system_1} takes form
\begin{equation*}\left\{
\begin{array}{lll}
u_{1x} & = &u_x \\
v_{1x} & = & v_x \, . \\
\end{array}\right.
\end{equation*}

\subsection{Case (2)}

Let us find  system \eqref{systemCartan}
possessing $n$-integrals \eqref{I_2} and \eqref{I_2^*}.
We compare the coefficients in $DI_2=I_2$ by $u_{xx}$ and $v_{xx}$ and get
\begin{equation}\label{ex5_*}
\begin{array}{lll}
2\tilde f_{u_x}+\tilde g_{u_x} &=&2   \\
2\tilde f_{v_x}+\tilde g_{v_x} &=&1.
\end{array}
\end{equation}
We also compare the coefficients in $DI_2^*=I_2^*$  and \\
$D(D_x^2I_2-2I^*_2)=(D_x^2I_2-2I^*_2)$ by $u_{xxxx}$ and $v_{xxxx}$ respectively and get $\tilde f_{u_x}=1$ and $\tilde g_{v_x} =1$.
It follows from \eqref{ex5_*} that  $\tilde f_{v_x}=0$ and $\tilde g_{u_x} =0$. Therefore,  our system \eqref{systemCartan} becomes
\begin{equation*}\left\{
\begin{array}{lll}
u_{1x} &=& u_x+  f(u,v,u_1,v_1)   \\
v_{1x} &=& v_x+ g (u,v,u_1,v_1).
\end{array}\right.
\end{equation*}
We write equality $DI_2=I_2$ and get
\begin{multline*}
2u_{xx}+2f_uu_x+2f_vv_x+2f_{u_1}(u_x+f)+2f_{v_1}(v_x+g)+v_{xx}+g_uu_x+g_vv_x+g_{u_1}(u_x+f)\\
+g_{v_1}(v_x+g) -2(u_x+f)^2 +2(u_x+f)(v_x+g)-(v_x+g)^2
=2u_{xx}+v_{xx}-2u_x^2+2u_xv_x-v_x^2\, .
\end{multline*}
By comparing the coefficients by $u_x$, $v_x$ and $u^0_xv^0_x$ in the last equality we obtain the system of equations
\begin{equation*}
\begin{array}{lll}
2f_u+f_{u_1}+g_u+g_{u_1} -4f+2g & = & 0 \\
2f_v+2f_{v_1}+g_v + g_{v_1}+2f-2g & = & 0 \\
2ff_{u_1}+2gf_{v_1}+fg_{u_1}+gg_{v_1}-2f^2+2fg-g^2 & = & 0\, .
\end{array}
\end{equation*}
That suggests the following change of variables
\begin{equation*}
u=P, \,\, u_1-u=Q, \,\, v=S, \,\, v_1-v=T
\end{equation*}
to be made.
In new variables the system \eqref{systemCartan} becomes
\begin{equation}\label{ex5_new_system_1}\left\{
\begin{array}{lll}
Q_x & = & F(P,Q,S,T)  \\
T_x & = & G(P,Q,S,T)\, .
\end{array}\right.
\end{equation}
The comparison of coefficients in $DI_2=I_2$ by $P_x$, $S_x$ and $P^0_xS^0_x$ gives
\begin{equation}\label{ex5_system_2}
\begin{array}{lll}
-4F+2G+2F_P+G_P & = & 0 \\
2F-2G+2F_S+G_S & = & 0 \\
-2F^2+G(-G+2F_T+G_T)+ F(2G+2F_Q+G_Q) & = & 0\, .
\end{array}
\end{equation}
The coefficients   in $DI_2^*=I^*_2$ by $S_{xxx}$ and $P_{xxx}$ are compared and we obtain the following equalities
\begin{equation}\label{ex5_system_3}
\begin{array}{lll}
F+F_S & = & 0 \\
-2F+F_P & = & 0\, . \\
\end{array}
\end{equation}
It follows from \eqref{ex5_system_2} and \eqref{ex5_system_3} that $G_S =  2G $,
$G_P  =  -2G$,
$F_S  =  -F$ and
$F_P = 2F$.
Therefore, system \eqref{ex5_new_system_1} can be written as
\begin{equation*}\left\{
\begin{array}{lll}
Q_x & = & A(Q,T)e^{-S+2P}  \\
T_x & = & B(Q,T)e^{2S-2P}\, .
\end{array}\right.
\end{equation*}
We compare the coefficient in $DI^*_2=I^*_2$ by $S_{xx}$ and get
\begin{equation*}
3e^{4P-2S}A^2-3e^{4P-2S}AA_Q=0,
\end{equation*}
that is $A=A_Q$. Hence, $A(Q,T)=e^Q\tilde A(T)$. Now we compare the coefficient in $DI_2=I_2$ by $P^0_xS^0_x$ and get
\begin{equation}\label{ex5_system_4}
\tilde A+\tilde A_T=\frac{1}{2}e^{-4P+3S-Q}(B-B_T)-\frac{\tilde A}{2B}B_Q\, .
\end{equation}
Since functions $\tilde A(T)$ and $B(Q,T)$ do not depend on variable $P$, then it follows from \eqref{ex5_system_4} that $B=B_T$, that is $B=\tilde B(Q)e^T$.
Now \eqref{ex5_system_4} becomes
\begin{equation*}
-2\frac{\tilde A+\tilde A_T}{\tilde A} =\frac{\tilde B_Q}{\tilde B}\, .
\end{equation*}
Note that the right side of the last equality depends on $Q$ only, while the left side depends on $T$ only. Hence,  $-2\frac{\tilde A+\tilde A_T}{\tilde A}=c $ and  $\frac{\tilde B_Q}{\tilde B}=c$,
where $c$  is some constant.  One can see that $\tilde A=c_1e^{-(2c+1)T}$ and $\tilde B=c_2e^{cQ}$ and therefore system \eqref{ex5_new_system_1} becomes
\begin{equation*}\left\{
\begin{array}{lll}
Q_x & = & c_1e^{-S+2P+Q-(2c+1)T}  \\
T_x & = & c_2e^{2S-2P+T+cQ},
\end{array}\right.
\end{equation*}
where $c$, $c_1$ and $c_2$ are some constants. Equality $DI_2-I_2=0$ becomes $-3cc_1c_2e^{s+(c+1)Q-2cT}=0$, which  implies that either $c=0$, or $c_1=0$, or $c_2=0$.
 Note that the  $DI_2^*=I_2^*$ is also  satisfied if either $c=0$ or $c_1=0$ or $c_2=0$. So we have three cases:

when $c=0$ the system \eqref{systemCartan} becomes \eqref{Example5_1} with $c_1=A$ and $c_2=B$.

when $c_1=0$ the system \eqref{systemCartan} becomes \eqref{Example5_2} with $c_2=B$.

when $c_2=0$ the system \eqref{systemCartan} becomes \eqref{Example5_3} with $c_1=A$.

\subsection{Case (3)}

Let us find  system \eqref{systemCartan}
possessing $n$-integrals \eqref{I_3} and \eqref{I_3^*}.
We compare the coefficients in $DI_3=I_3$ by $u_{xx}$ and $v_{xx}$ and get
\begin{equation}\label{ex6_*}
\begin{array}{lll}
\tilde f_{u_x}+\frac{1}{3}\tilde g_{u_x} &=&1   \\
\tilde f_{v_x}+\frac{1}{3}\tilde g_{v_x} &=&1.
\end{array}
\end{equation}
We also compare the coefficients in $DI_3^*=I_3^*$  and
$D(D_x^4I_3-I^*_3)=(D_x^4I_3-I^*_3)$ by $u_{(6)}$ and $v_{(6)}$ respectively and get $\tilde f_{u_x}=1$ and $\tilde g_{v_x} =1$.
It follows from \eqref{ex6_*} that  $\tilde f_{v_x}=0$ and $\tilde g_{u_x} =0$. Therefore,  our system \eqref{systemCartan} becomes
\begin{equation*}\left\{
\begin{array}{lll}
u_{1x} &=& u_x+  f(u,v,u_1,v_1)   \\
v_{1x} &=& v_x+ g (u,v,u_1,v_1).
\end{array}\right.
\end{equation*}
By comparing the coefficients by $u_x$, $v_x$ and $u^0_xv^0_x$ in $DI_3=I_3$ we obtain the system of equations
\begin{equation*}
\begin{array}{lll}
f_u+f_{u_1}+\frac{1}{3}g_u+\frac{1}{3}g_{u_1} -2f+g & = & 0 \\
f_v+f_{v_1}+\frac{1}{3}g_v + \frac{1}{3}g_{v_1}+f-\frac{2}{3}g & = & 0 \\
ff_{u_1}+gf_{v_1}+\frac{1}{3}fg_{u_1}+\frac{1}{3}gg_{v_1}-f^2+fg-\frac{1}{3}g^2 & = & 0\, .
\end{array}
\end{equation*}
That suggests the following change of variables
\begin{equation*}
u=P, \,\, u_1-u=Q, \,\, v=S, \,\, v_1-v=T
\end{equation*}
to be made.
In new variables the system \eqref{systemCartan} becomes
\begin{equation}\label{ex6_new_system_1}\left\{
\begin{array}{lll}
Q_x & = & F(P,Q,S,T)  \\
T_x & = & G(P,Q,S,T)\, .
\end{array}\right.
\end{equation}
The comparison of coefficients in $DI_3=I_3$ by $P_x$, $S_x$ and $P^0_xS^0_x$ gives
\begin{equation}\label{ex6_system_**}
\begin{array}{lll}
6F - 3G - 3F_P - G_P & = & 0 \\
-3F + 2G - 3F_S - G_S & = & 0 \\
F^2 - FG + \frac{1}{3}G^2 - 2GF_T - \frac{1}{3}GG_T - FF_Q -\frac{1}{3}FG_Q & = & 0\, .
\end{array}
\end{equation}
The comparison of coefficients in  $DI_3^*=I^*_3$ by $S_{(5)}$ and $P_{(5)}$  gives
\begin{equation}
\begin{array}{lll}\label{ex6_system_**+}
F+F_S & = & 0 \\
-2F+F_P & = & 0\, . \\
\end{array}
\end{equation}
Using equations \eqref{ex6_system_**} and \eqref{ex6_system_**+} we get
$G_S  =  2G$,
$G_P  =  -3G $,
$F_S  =  -F$, and
$ F_P =  2F$.
Therefore, system \eqref{ex6_new_system_1} can be written as
\begin{equation*}\left\{
\begin{array}{lll}
Q_x & = & A(Q,T)e^{-S+2P}  \\
T_x & = & B(Q,T)e^{2S-3P}\, ,
\end{array}\right.
\end{equation*}
where $A$ and $B$ are some functions depending on $Q$ and $T$ only. We compare the coefficients in $DI_3-I_3=0$ by $S_x^0P_x^0$ and the coefficients in $DI_3^*-I_3^*=0$ by $P_{(4)}$, $S_{(4)}$ and $P_{(3)}P_x$ respectively and get
\begin{equation}\label{ex6_system_***}
\begin{array}{lll}
a_{11}A_T + a_{12}B_T + a_{13}A_Q + a_{14}B_Q + b_1 & = & 0 \\
a_{21}A_T + a_{22}B_T + a_{23}A_Q + a_{24}B_Q + b_2 & = & 0 \\
a_{31}A_T + a_{32}B_T + a_{33}A_Q + a_{34}B_Q + b_3 & = & 0 \\
a_{41}A_T + a_{42}B_T + a_{43}A_Q + a_{44}B_Q + b_4 & = & 0 ,\\
\end{array}
\end{equation}
where
\begin{equation*}
\begin{array}{llll}
a_{11}=-e^{-P+S}B, & a_{12}=-\frac{1}{3}e^{-6P+4S}B, & a_{13}=-e^{4P-2S}A, &  a_{14}=-\frac{1}{3}e^{-P+S}A,\\
a_{21}=-33e^{-P+S}B, & a_{22}=-11e^{-6P+4S}B, & a_{23}=-28e^{4P-2S}A, &  a_{24}=-11e^{-P+S}A,\\
a_{31}=-13e^{-P+S}B, & a_{32}=-\frac{13}{3}e^{-6P+4S}B, & a_{33}=-16e^{4P-2S}A, &  a_{34}=-\frac{13}{3}e^{-P+S}A,\\
a_{41}=18e^{-P+S}B, & a_{42}=-79e^{-6P+4S}B, & a_{43}=328e^{4P-2S}A, &  a_{44}=6e^{-P+S}A,\\
\end{array}
\end{equation*}
and
\begin{equation*}
\begin{array}{l}
b_1=e^{4P-2S}A^2-e^{-P+S}AB+\frac{1}{3}e^{-6P+4S}B^2  \\
 b_2=28e^{4P-2S}A^2-33e^{-P+S}AB+11e^{-6P+4S}B^2 \\
b_3=16e^{4P-2S}A^2-13e^{-P+S}AB+\frac{13}{3}e^{-6P+4S}B^2 \\
 b_4=-328e^{4P-2S}A^2+18e^{-P+S}AB+79e^{-6P+4S}B^2\, .\\
\end{array}
\end{equation*}
We solve the linear system of equations \eqref{ex6_system_***} with respect to $A_T$, $A_Q$, $B_T$ and $B_Q$ and get the following system of differential equations
$A_T = -A$,
$A_Q  =  A $,
$B_T  =  B $ and
$B_Q  =  0 $.
Thus the system \eqref{ex6_new_system_1} is written as
\begin{equation*}\left\{
\begin{array}{lll}
Q_x & = & c_1e^{2P+Q-S-T}  \\
T_x & = & c_2e^{-3P+2S+T}\, ,
\end{array}\right.
\end{equation*}
where $c_1$ and $c_2$ are arbitrary constants. It is equivalent to system \eqref{Example6} with $A=c_1$ and $B=c_2$.


\begin{thebibliography}{100}



\bibitem{ZhMHSh} A.V. Zhiber, R.D. Murtazina, I.T. Habibullin,  and  A.B. Shabat, Characteristic Lie rings and integrable models in mathematical physics, \textit{Ufa Math. J.} \textbf{4} (3)(2012) 17--85.

\bibitem{HZhS} I.T. Habibullin, N. Zheltukhina,  and A. Sakieva,  Discretization of hyperbolic type Darboux integrable equations preserving integrability, J. Math. Phys. \textbf {52}  (2011) 093507--093519.

\bibitem{HZh} I.T. Habibullin  and N. Zheltukhina, Discretization of Liouville type nonautonomous equations \textit{J. Nonlinear Math. Phys.} \textbf {23}  (2016) 620--642.

\bibitem{ZhZh}  K. Zheltukhin and N. Zheltukhina,   On the discretization of Laine equations, \textit{ J. Nonlinear Math. Phys.} \textbf {25}  (2018) 166--177 .

\bibitem{HP} I.T. Habibullin, A. Pekcan,  Characteristic Lie algebra and the classification of semi-discrete models, \textit{Theoret. and Math. Phys.} \textbf {151} (2007) 781–790.

\bibitem{KoZh} O.S. Kostrigina and A.V. Zhiber,   Darboux-integrable two-component nonlinear hyperbolic systems of equations, \textit{ J. Math. Phys.} \textbf {52} (2011) 033503--033535.

\bibitem{ShY} A.B. Shabat  and R.I. Yamilov   Exponential Systems of Type I and the Cartan Matrices (Russian) \textit {Preprint BBAS USSR~Ufa} (1981).

\bibitem{I} N.H. Ibragimov, A.V. Aksenov, V.A. Baikov, V.A. Chugunov, R.K. Gazizov and A.G. Meshkov
 CRC Handbook of Lie Group Analysis of Differential Equations  Vol. 2. Applications in Engineeringand Physical Science, edited by Ibragimov \textit{ Boca Raton, FL: CRC Press} (1995).

\bibitem{GT}  E.I. Ganzha and S.P. Tsarev, Integration of Classical Series An, Bn, Cn, of Exponential Systems \textit {Krasnoyarsk: Krasnoyarsk State Pedagogical University Press} (2001).

\bibitem{LS} A.N. Leznov and M.V. Savel'ev,  Group Methods of Integration of Nonlinear Dynamical Systems, \textit {Progress in Physics} \textbf {15} \textit {Birkhäuser Verlag, Basel} (1992).

\bibitem{HZhY} I.T. Habibullin, K. Zheltukhin, and M. Yangubaeva,  Cartan matrices and integrable lattice Toda field equations, \textit {J. Phys. A} \textbf {44} (2011) 465202--465222.

\end{thebibliography}
\end{document}